\def\ga{\mathrel{\mathpalette\fun >}}
\def\fun#1#2{\lower3.6pt\vbox{\baselineskip0pt\lineskip.9pt
  \ialign{$\mathsurround=0pt#1\hfil##\hfil$\crcr#2\crcr\sim\crcr}}}
\relax
\documentstyle[12pt]{article}
\advance\textheight by \topskip
\begin{document}
\thispagestyle{empty}
\begin{flushright}
SU-ITP-93-20\\
hep-ph/9308218\\
\today
\end{flushright}
\vskip 1.2cm
\begin{center}
{\Large\bf Alignment in Gamma-Hadron Families\\
\vskip .7cm
Detected by Deep Lead X-ray Emulsion Chambers}\\
\vskip 1.4cm
\vskip 1.2cm
{\bf I.P. Ivanenko$^1$, V.V. Kopenkin$^1$, A.K. Managadze$^1$,
               and   I.V. Rakobolskaya$^{1,2}$}
\vskip 0.8cm
$^1$ Institute of Nuclear Physics,
   Moscow State University, Moscow 119899, Russia
\vskip 0.05cm
$^2$  Department of Physics, Stanford University, Stanford
CA 94305, USA
\vskip 0.5cm
\end{center}
\hskip 2cm

{\centerline{\large ABSTRACT}}
\begin{quotation}
Alignment  of main fluxes of  energy in a target plane is found in
families of cosmic ray particles detected in deep lead X-ray emulsion
chambers. The fraction of aligned events becomes unexpectedly large
for superfamilies with the total energy of gamma-quanta exceeding
$10^3$ TeV. This can be considered as an evidence of
existence of coplanar scattering of secondary particles in
interactions of primary particles with  superhigh energies ($E_0 \ga
10^{16}$ eV).\\

Data analysis suggests that production of most  superfamilies
happens low above the chamber and is characterized by a coplanar
scattering and quasiscaling spectrum of secondary
particles in the fragmentation region.

 \end{quotation}
 \newpage

\section{Introduction}

International collaboration ``Pamir'' is conducting a cosmic ray
experiment  at the altitude 4400 meters in Pamir mountains.  Our
equipment consists of X-ray emulsion  chambers of two kinds:
C-chambers and Pb-chambers. C-chambers consist of   a block of carbon
 covered  from both sides by blocks of lead of thickness 6 cm. Each
block of lead contains  3 layers of X-ray film. Pb-chambers are
assembled from many carpets of lead of thickness 1 cm, interlaid
with X-ray film. The total depth of  each chamber usually is 60 cm.
Once a year all these chambers are disassembled, the X-ray  film is
taken away and the results of the experiment are investigated.
The new results to be reported in this paper were obtained using
Pb-chambers.

Gamma quanta and hadrons which are present in the cosmic rays create
electron-photon cascades in the lead. These cascades are registered
in the X-ray film as dark spots of a size which is typically smaller
than 1 mm.

The darkness density $D(E,t) $ of each spot depends on energy $E$ of
the cascade and of the depth $t$ of its development in the chamber.
By comparison of the distribution of $D(E,t) $ for all spots with the
theoretical predictions one can obtain the energy of each cascade
and, consequently, the energy of the primary particles producing
these cascades. We register particles with energies beginning from
2-6 TeV. \, Gamma quanta produce electron-photon cascades in the
upper part of the chamber only, whereas  hadrons produce such
cascades at large depth as well. Efficiency of registration of
hadrons is about 70\%. Such a deep lead chamber works very similar to
an ionization calorimeter.

Apart from single particles, this chamber registers also the
so-called ``gamma-hadron families". Each such family is a
genealogically related group of gamma quanta, electrons and hadrons,
which appear as a result of development of a nuclear electromagnetic
shower created when the primary high energy cosmic rays enter the
atmosphere.

By measuring of the coordinates and directions of motion of particles
inside the chamber one can reconstruct the target diagram of the
family and find its properties, such as the total energy of gamma
quanta in the family $\sum{E_\gamma}$, the distribution of  gamma
quanta and hadrons in the area, the energy released by hadrons to the
electron-photon cascades $E_h^\gamma$, etc. All families in our
experiment were  classified by the the value of total energy of gamma
quanta  $\sum{E_\gamma}$. We considered families with $\sum{E_\gamma}
> 100$ TeV. When we studied ``superfamilies'' with  $\sum{E_\gamma}
\ga 1000$ TeV, we have found that near the center of the event at
the X-ray film one can often see one or more large diffuse dark
spots, of a size from few millimeters to several centimeters. Each
such halo usually appears as a result of development of an
electron-photon cascade   by a high energy gamma  quanta  at some
height above the chamber \cite{A,B}.

In a lower part of a deep lead chamber one can find also large spots
which also look like a halos, but have   hadronic origin  \cite{C}.
Each such halo is a result of a cascade produced by hadrons of very
high energy (about $100 \div 500$ TeV) in lead.

In 1985 Pamir Collaboration has found several families with 3 or 4
halos of electromagnetic origin   \cite{D,E}, and in most of these
families (in 5 out of 6 such families) these halos were aligned.
Experimental results obtained during the subsequent years did not
increase considerably statistics for  investigation of these events,
but the relative fraction of events with the aligned halos of
electromagnetic origin became smaller.

As an alignment parameter we used  the function
\begin{equation}\label{1}
\lambda_m = \sum_{i\neq j\neq k}^{m}{{\cos 2\phi_{ijk}} \over
m(m-1)(m-2)} \ ,
\end{equation}
where $m$ is the number of halos and $\phi_{ijk}$ is the angle
between the two vectors, which connect the center of the halo $k$
with the centers of the halos $i$ and $j$ \cite{F}. An event is
considered aligned if $\lambda \geq 0.6$. (A stronger requirement is
$\lambda \geq 0.8$.)

Parameter $\lambda$  is the best known parameter of asymmetry
describing the  degree of alignment rather than  eccentricity. For
example, $\lambda_4$ will be large if four points belong to the same
straight line, but it will be small if  these points form four
vertices of a long rectangle.

We performed a computer simulation of families with several halos
using quasiscaling models without inventing any  specific mechanisms
for producing asymmetry \cite{G,I,J}. Relative fraction of events
with three aligned halos in the simulated families was rather high,
about $30\div35\%$. This number gives the level of background noise,
i.e. the level of of fluctuations in the development of the
nuclear-electromagnetic cascade. According to simulations,
the degree  of alignment for three random `particles', which do not
belong to the same cascade, was given by $24\%$.

However, in the works discussed above we registered only halos at the
same (small) depth in the upper part of C-chambers, under some
constraint on the level of darkness of the spots $D$ on the X-ray
film.

Experimental results obtained in Pb-chambers  allowed us to
investigate alignment of halos at different depths and with
different levels of darkness $D$, and to take into account the
contribution of hadronic  cascades  in the lower part of a chamber
\cite{G}.  We have found that the alignment of halos in the same
family is a function of depth and of the level of darkness.

In our investigation of alignment we tried to find a better method of
selection of objects to be examined, which would be  more sensitive
and less dependent on methodological factors. In a search for such
method we proposed to study not only halos, but a more general class
of objects, which we called `Energy Distinguished Cores'  (EDC)
\cite{I,K}. These objects on the X-ray film
correspond to the centers of the most  jets with the highest energy
in a family.
They include the following objects:
\begin{itemize}
\item halos of the electromagnetic origin (or separate centers of one
halo)
\item gamma-clusters (i.e. compact groups of gamma-quanta which are
combined into separate clusters using the criterions of
(compact
decascading)
or close,
\item separate gamma-quanta of very high energy
perhaps)
\item high energy hadrons, in particular the hadrons which produced
halos in the chamber
\end{itemize}
In order to treat gamma-quanta and hadrons in a similar way, one
should multiply by the factor of $3$ the energy $E^{(\gamma)}_h$
released by a hadron in the chamber into the electromagnetic
component, since in average $<k_\gamma> = 0.33$.

This approach allows to analyse  alignment in the gamma-hadron
families of not very high  energies  which do not contain halos, and
to avoid discrimination of some  types of EDC against some other
ones. By this method we effectively investigate spatial distribution
of the most energetic particles in the shower, where the distribution
of hadrons correspond to the distribution of charged particles in the
shower, and the distribution of gamma-quanta correspond to neutral
secondary particles.

To investigate alignment of all events which occur at different depth
in the chamber, we made the target diagram: we projected all events
onto one plane by shifting their images along the direction of the
shower. After that we studied alignment of  all EDC in this plane.

For example, we were looking for alignment of four EDC. In one of the
superfamilies  there were four electromagnetic halos which were not
aligned. However, in the lower part of the chamber we have found two
hadrons which had energy higher than the energy of the two halos out
of four. After we projected the halos and the hadrons onto the same
plane, we have found that the four most energetic events  (two halos
and two hadrons) were aligned in accordance with the criterion
$\lambda_4 > 0.8$.

\section{Alignment of EDC}
To analyse the effect of alignment  we studied 74 gamma-hadron
families  with energies $100$ TeV $< \sum E_\gamma < 5000$ TeV, which
were found in Pb-chambers.
Experimental results were compared with the  simulation of  random
falling down of 3 and 4 points, and with the simulation of a family
described by the model of quasiscaling interaction \cite{J}.

The most significant deviation of experimental results  from the
results obtained by the simulations appears if one investigates
alignment of four EDC with the alignment  criterion $\lambda_4 >
0.8$. At $\sum E_\gamma \sim 100\div 300$ TeV the fraction of aligned
events was $(9\pm 3)\%$, at   $\sum E_\gamma  \sim 300\div 500$ TeV
this fraction was $(23\pm 10)\%$. The effect grows with energy, and
at $\sum E_\gamma  \sim 500\div 5000$ approaches $(43\pm 17)\%$.
Meanwhile according to our simulations of families this effect does
not depend on energy and is equal to $8\%$, whereas the simulation of
randomly falling uncorrelated particles gives only $4\%$.

Analogous analysis was performed in \cite{L} for other  criterions of
alignment as well, for example, for $\lambda_3 > 0.6$ and $\lambda_3
> 0.8$ for three EDC, and for $\lambda_4 > 0.6$ for four EDC. In
addition, an investigation of alignment was performed for different
components of families: 4 hadrons, 4 gamma-clusters, 4 gamma-quanta.
All results are in a good correlation with each other. They show
increase of the fraction of aligned events when one goes from
gamma-quanta to gamma-clusters, to hadrons, and then finally to EDC.

Statistical analysis with the use of various criterions (rank
correlation of Spearman, Kolmogorov-Smirnov criterion) has
shown that the experimental  results on alignment of EDC at $\sum
E_\gamma > 500$ TeV can be distinguished from the  background
fluctuations in our simulations at $95\%$ confidence level. At the
same  confidence level, alignment  grows with energy $\sum E_\gamma$.

Conditions of registration of particles influence the results because
of the limited accuracy of energy determination, and also because of
possible deviations of the coefficient   $k_\gamma$  from its average
value $\langle k_\gamma\rangle = 0.33$. Simulation of different
possibilities has shown that an account taken of these effects may
increase almost twice the fraction of  events in our experimental
data which we interpret as aligned. This suggests that the fraction
of aligned EDC may be even greater than we thought.

The analysis was made also for every family selecting various number
of EDC from 3 to 10.  Results for families with small energies come
close to the fluctuation level of model simulations. On the other
hand, the fraction of aligned events for 14 superfamilies with $\sum
E_\gamma > 500$ TeV  exceeds by two standard deviations the
fluctuation level of model simulations for any particular number of
EDC. One should  note, however, that statistics is very small. We
have found 6 events with 4 aligned cores, 4 events with 5 aligned
cores, 2 events with 6 aligned cores and one event with seven
aligned cores.

14 superfamilies with $\sum E_\gamma > 500$ TeV
were divided into groups with or without alignment.  For these two
groups various characteristics were analysed. No statistically
significant differences  between these two classes of families were
found.

The energy distribution for 4  centers of highest energy in
superfamilies suggests that we are observing the fragmentation part
of the spectrum of created particles, which is not considerably
modified by the cascade and by the processes inside the chamber. It
suggests also that the highest energy objects of most of the
superfamilies were created in a single act of interaction at a
relatively small altitude, probably about 2 km above a chamber
\cite{M}.

General analysis of the nuclear-electromagnetic cascades, supported
by model simulations of \cite{N}, shows that alignment should appear
due to the interaction with pronounced coplanarity not far from
a chamber, since  development of the nuclear-electromagnetic
cascade ``blurs'' the alignment after few interactions.

Assuming the most probable interaction altitude $H = 2$ km and
measuring the distance between aligned cores, the momentum $p_t$
transferred between a core and the center of a 4 core jet ( or group)
is estimated as $p_t \sim 1$ GeV.
A typical momentum of a core $p_t^t$ transverse to the axis of
alignment is 10 times less, i.e. $<p_t > \sim 0.1$ GeV.

Average invariant mass $M$ of the entire  subfamily of 4 aligned
particles is $\langle M^2\rangle  =(60^{+120}_{-60})$ GeV$^2$.

  It is difficult to explain    coplanarity of secondaries   within
conventional models of  interaction.   Muhamedshin and  Slavatinsky
\cite{N} have shown   that the magnetic field of the Earth could not
be  responsible  for  any appreciable asymmetry.  I. Royzen \cite{O}
has suggested to interpret the phenomenon of alignment as a
projection of rupture of the quark-gluon string produced in the
process of hard double inelastic diffraction dissociation, the string
being inclined between a semihard scattered fast quark and the
incident hadron remnants. Such explanation seems plausible because
the energy threshold of the alignment effect is suitable for hard
double inelastic diffraction. In this case a target diagram of a
superfamily with alignment may be considered as a direct
`photographic image of such process.

It would be most desirable to test this effect on accelerators.
Preliminary estimates indicate that the energies accessible at FNAL
would be barely enough to produce families with alignment. However,
one could still obtain interesting results at these energies due to
the possibility to obtain much better statistics that in cosmic rays.

We would like to express our gratitude   E.L. Feinberg,
T.M. Roganova,  I.I Royzen, S.A. Slavatinsky and G.T. Zatsepin for
active discussion of the results and E.G. Popova, E.I. Pomelova and
N.G. Zelevinskaya for the large work they did participating in this
experiment. We are very grateful to J. Bjorken for a very interesting
discussion of the possibility of related accelerator experiments. One
of the authors (I.R.) is greatly thankful to the Department of
Physics of Stanford University, and especially to R. Wagoner for
their kind hospitality  at Stanford, where this paper was completed.

\end{document}